\documentclass[MSNbibl,nameyear,dvips]{arxstspdf}
\usepackage{flushend}
\usepackage{stfloats}

\volume{26}
\issue{2}
\pubyear{2011}
\firstpage{231}
\lastpage{234}
\doi{10.1214/11-STS331B}
\referstodoi{10.1214/10-STS331}

\begin{document}
\begin{frontmatter}

\title{Discussion of ``Bayesian Models and Methods in
Public Policy and Government Settings'' by S. E. Fienberg}
\runtitle{Discussion}
\pdftitle{Discussion of Bayesian Models and Methods in
Public Policy and Government Settings by S. E. Fienberg}

\begin{aug}
\author{\fnms{Graham} \snm{Kalton}\corref{}\ead[label=e1]{grahamkalton@westat.com}}
\runauthor{G. Kalton}

\affiliation{Westat}

\address{Graham Kalton is Senior Vice President, Westat, 1600
Research Blvd, Rockville, Maryland 20850, USA \printead{e1}.}

\end{aug}

% ABSTRACT

% KEYWORDS

\end{frontmatter}

Steve Fienberg has presented a wide and interesting range of
applications of Bayesian methods in public policy and government
settings (including election night forecasting which I might prefer to
classify as fleeting public entertainment!). The examples exhibit the
common feature that they all involve highly complex problems that are
difficult to handle in a non-Bayesian framework. Sedransk (\citeyear{Sedransk08})
has provided some other examples of Bayesian methods in such settings which
also share this feature. I am sympathetic to the use of Bayesian
methods in such special circumstances, as illustrated below.

My initial comments focus on the choice of modes of inference for
large-scale government surveys, particularly surveys of households and
persons, that are the backbone for satisfying policy and government
data needs. An important feature of these surveys, in common with most
surveys, is that they typically collect data on many variables and
these data are then used to produce very large numbers of estimates. In
this area, I generally favor the frequentist repeated sampling mode of
inference, commonly termed design-based inference (Kalton, \citeyear{KAL02}), and I
believe that my views are in line with most other survey statisticians
(see, e.g., Rao, \citeyear{RAO}, in this issue). However, there are situations in
which design-based inference cannot satisfy analytic objectives. Also,
limitations in the practical application of de\-sign-based inference are
becoming increasingly troublesome. To the extent possible, I prefer to
minimize the dependence of survey estimates on statistical models. When
models are needed, I prefer non-Bayesian models to Bayesian models, but
I accept that Bayesian models have major analytic attractions for some
complex analytic problems. My chosen focus excludes discussion of
applications of what are often termed ``the analytic uses of survey
data.'' For example, when a survey collects data for a non-randomized
observational study, models are clearly essential to evaluate the
effects of different levels of program exposure; this kind of modeling
is outside my current scope.

To start, consider the ideal situation of a survey that uses a sampling
frame with complete coverage of the finite target population, that
achieves complete response from all sampled elements, and that has a
sample size chosen to be large enough to produce design-based estimates
of adequate precision for prespecified policy needs. In such a case,
the design-based approach has major attractions for a~typical survey,
especially in view of the multipurpose nature of surveys which aim to
produce a~multitude of descriptive estimates. Under this mode of
inference, the survey estimates are not model-depen\-dent. To expand on
George Box's often quoted saying ``All models are wrong, but some are
useful,'' I~would add the caution for the survey context that ``Models
are not always useful.'' Models need to be carefully developed and
tested if model-dependent inference is to be used, particularly with
large-scale surveys. With a small sample, a model-dependent estimate
may be preferred because its mean squared error (MSE) is less than the
large variance of the design-based estimate; however, with a large
sample, the bias associated with the model-dependent estimate becomes
the dominant factor in the MSE. Besides the precision of the estimates,
another important attribute of quality in government statistics is the
timeliness with which the estimates are produced. All the many
design-based estimates from a~survey can be produced relatively quickly
since they do not require the time needed to develop and test many
models. Also, the design-based approach has the flexibility of readily
permitting the computation of additional estimates if the initial
findings indicate they may be of interest.

Although design-based estimates are not dependent on the validity of
statistical models, models do play important roles in survey sample
design and analysis. Implicit and explicit models have been involved in
sample design since the early days including, for instance, in
stratification and the choice of the clusters at various stages of
sampling (see Sedransk, \citeyear{Sedransk08}, for a~discussion of Bayesian models in
design). Also, models have long been used in analysis, through such
techniques as poststratification, and ratio and regression estimation.
The distinctive feature of the use of such models in design-based
inference is that the sample estimates are approximately unbiased
irrespective of the suitability of the models. The choice of model
affects only the precision of the survey estimates. S{\"a}rndal, Swensson and
Wretman (\citeyear{SarSweWre92}) conveyed this meaning by titling their book ``Model Assisted
Survey Sampling.'' The aim of sample design is to develop efficient
model-assisted estimates in order to avoid the need for model-dependent
estimates.

The ideal situation described above is unfortunately seldom encountered
in survey practice: the sampling frame rarely provides complete
coverage of the target population and nonresponse---both unit (total)
and item nonresponse---is almost inevitable when survey data are
collected from the public. Indeed, a major current concern in survey
research is the continuing decline in response rates. Also, with
landline telephone surveys, the noncoverage rate is increasing as more
households are relying only on cell phones. Such sample deficiencies
are a limitation for design-based inference. Nonresponse and
noncoverage weighting adjustments are used to attempt to reduce the
biases in survey estimates resulting from unit nonresponse and
noncoverage and imputation methods are widely used to assign values for
item nonresponses (Brick and Kalton, \citeyear{BriKal1996}). Such weighting adjustments
and imputations are necessarily model-dependent (as is the approach
that simply analyzes the reported data). Thus even with the
design-based approach, some reliance on models is inevitable. The aim
is to limit the dependence of the survey estimates on models by
minimizing the impact of missing data.

The possible models for use in weighting adjustments are generally
constrained by the limited set of auxiliary variables available (data
for both respondents and nonrespondents for nonresponse adjustments and
exactly comparable data for respondents and the target population for
noncoverage adjustments). The effect of the weighting adjustments on
the variances of survey estimates can be readily captured using
replication methods that include a replication of the adjustments. Most
imputation procedures are based on non-Bayesian regression-type models,
using responses to other items in the survey to predict the missing
responses. Including the effect of imputation on variances is less
straightforward, but a range of methods have been proposed within the
design-based framework (e.g., Fay, \citeyear{FAY91}; Rao and Shao, \citeyear{RaoSha92};
S\"{a}rndal, \citeyear{SAR}; Shao and Steel, \citeyear{ShaSte99}; Kim and Fuller, \citeyear{KimFul04}; Haziza
and Rao, \citeyear{HAZRAO}). Bayesian methods have been applied for imputation,
particularly using multiple imputation methods (see Schenker et al.,
\citeyear{Schetal11}, for a recent example and references to many earlier
applications). An example of the application of a~highly complex
Bayesian hierarchical multiple imputation model is described by
Heeringa, Little and Ragunathan (\citeyear{HEELITAND02}); this example involves a multivariate model of
components of wealth for use when some respondents cannot report exact
amounts for some of the components but they can often provide brackets
within which the amounts lie. While multiple imputation provides a~means of
taking imputation variance into account, it is not a~panacea.
It provides consistent variance estimates only for certain estimates
for which allowance is made in the imputation model construction (Kim
et al., \citeyear{Kimetal06}). It does not, for instance, provide consistent variance
estimates for unplanned domain estimates.

As Steve discusses, an area where the design-based approach clearly
fails is that of small area estimation. In the past few decades policy
makers have been increasingly demanding survey estimates for local
areas so that they can target their programs more effectively. Yet, it
is impractical to have survey sample sizes large enough to support
estimates for such local areas as U.S. counties or school districts
(and often even for states). Statistical models that use the survey
data together with related local area data as auxiliary information are
necessary to produce local area, model-dependent estimates. These
models, which ``borrow strength'' from other areas through the
auxiliary data, have been used for many years in U.S. federal programs
(see Schaible, \citeyear{SCH96}, for some examples) and their use is increasing.
Many applications employ non-Bayesian or empirical Bayes methods to
implement a hierarchical model, such as the Fay--Herriot model (which
can be viewed as either a standard mixed model or an empirical Bayes
model). These models satisfy many needs but there are situations where
the full Bayesian approach is advantageous. With area level modeling, a
Bayesian approach can be attractive when the sampling model does not
match the linking area level model (Rao, \citeyear{Rao03}). In such a case, a
Bayesian approach can take advantage of the powerful MCMC algorithm,
the software for which is readily available; however, the approach is
highly computer-intensive. See, for example, Mohadjer et al. (\citeyear{MOHetal}) for
an application of this approach, using WinBugs, for estimating adult
literacy in U.S. counties based on the National Assessment of Adult
Literacy survey. Another example of the application of hierarchical
Bayesian methods for small area estimation is the annual production of
state and substate estimates from the National Survey on Drug Use and
Health, started in 1999, with unit level mixed logistic and Poisson
models (Folsom, Shah and Vaish, \citeyear{FOLSHAVAI99}).

The development of small area models that are used, like those in U.S.
Census Bureau's Small Area Income and Poverty Estimates (SAIPE)
program, to allocate large amounts of government funding to local areas
is a time-consuming activity. Moreover, an extensive and thorough
testing program should be undertaken to assess the suitability of the
models (see, e.g., Citro, Cohen and Kalton, \citeyear{CITCOHKAL98}, for
a~detailed evaluation of the
1993 SAIPE county estimates of school-age children in poverty). When
Bayesian models are used, in addition to other model testing, I think
that the analyst should carefully examine and document how sensitive
the small area estimates are to the choice of the prior distributions.

The success of small area estimation models depends ultimately on the
availability and appropriate use of effective auxiliary variables in
the models. For example, in times when changes are occurring, some of
the relevant auxiliary variables need to be up-to-date, for otherwise
the estimates will be distorted. Before embarking on a small area model
approach to serve the needs of a major policy study, a careful
appraisal should be conducted to determine whether appreciable biases
could occur because of lack of important auxiliary data.

An area of current development extends the small area modeling to
encompass data collected in other, larger, surveys. If the larger
survey provides estimates for the variables of interest for small area
modeling that are sufficiently close to those produced by the original
survey, the dependent variables in the small area modeling may simply
be changed to those derived from the larger survey, as is the case with
the replacement of poverty estimates from the Current Population Survey
by the corresponding estimates from the much larger American Community
Survey in the SAIPE program (Bell et al., \citeyear{BELetal}). However, the estimates
from the larger survey are often not sufficiently close: they may be of
lower quality, perhaps using a different mode of data collection, they
may not cover exactly the same survey population, and the variables may
not be exactly comparable. While design-based methods may be available
for some cases involving combinations of surveys (e.g., Kim and Rao,
\citeyear{KIMRAO}), a Bayesian approach for combining data from several sources will
often be attractive in complex situations. As an example, to produce
county-level estimates of smoking and mammography screening rates,
Raghunathan et al. (\citeyear{Ragetal07}) employed a hierarchical Bayesian modeling
approach that combined data from three sources: the National Health
Interview Survey (NHIS), conducted by face-to-face interviewing; the
much larger Behavioral Risk Factor Surveillance System (BRFSS),
conducted by telephone only in households with landline phones; and
county-level covariates. The complex multivariate model with three
dependent variables (estimates for persons in NHIS households with
landline telephones, in NHIS households without landline telephones,
and in BRFSS households) is well suited for the use of the MCMC
technique of Gibbs sampling. Since the combination of data from several
surveys and administrative records can serve a number of policy
purposes (Schenker and Raghunathan, \citeyear{SchRag07}), the use of combinations of this type
is likely to expand considerably in the future. Combining data in this
way will often be facilitated by the analytic tools available in
Bayesian analysis. Model validation of such complex models requires
careful attention.

In summary, I believe that, despite the limitations noted earlier, the
design-based mode of inference should remain the main mode of inference
for descriptive estimates from large-scale government surveys. However,
model-dependent approaches are appropriate in circumstances such as
small area estimation where design-based inference cannot produce the
required estimates with adequate precision, and sometimes in the
developing field of combining data from surveys and other data sources.
In general, I favor non-Bayesian models, but there are cases where a
non-Bayesian approach is either extremely difficult or not workable. I~accept the use of Bayesian models in situations where their powerful
computing methods are needed, with the additional proviso that the
robustness of the model estimates to the choice of the prior
distributions should be carefully assessed.

% imsref loaded by svajune.rapalyte, 2011-05-06 10:34:59

\end{document}